\documentclass[12pt,epsf,amssymb]{article}
\usepackage{tabularx}
\usepackage{pstricks}
\usepackage{array}
\usepackage{graphics}
\usepackage{graphicx}
\usepackage{epsfig}
\usepackage{amsmath}
\usepackage{amssymb}
\usepackage{citesort}
\usepackage{axodraw}
\usepackage{feynmf}

\makeatletter

\usepackage{verbatim}

\newcommand{\bea}{\begin{eqnarray}}
\newcommand{\eea}{\end{eqnarray}}
\newcommand{\beq}{\begin{equation}}
\newcommand{\eeq}{\end{equation}}

\newcommand{\pdir}{p\kern -5.2pt\raise 0.2ex\hbox {/}}
\newcommand{\vdir}{v\kern -5.75pt\raise 0.15ex\hbox {/}}
\newcommand{\kdir}{k\kern -5.75pt\raise 0.15ex\hbox {/}}
\newcommand{\epsdir}{\epsilon\kern -5.0pt\raise 0.15ex\hbox {/}}
\newcommand{\bvdir}{\bar{v}\kern -5.75pt\raise 0.15ex\hbox {/}}
\newcommand{\Ddir}{D\kern -7.75pt\raise 0.20ex\hbox {/}}
\newcommand{\Adir}{A\kern -7.75pt\raise 0.20ex\hbox {/}}
\newcommand{\ldir}{l\kern -5.0pt\raise 0.2ex\hbox{/}}
\newcommand{\varepsdir}{\varepsilon\kern -5.5pt\raise 0.15ex\hbox{/}}

\newcommand{\taud}{\tau_{\frac{1}{2}}}
\newcommand{\taut}{\tau_{\frac{3}{2}}}

\newcommand{\lgl}{\langle}
\newcommand{\rgl}{\rangle}

\def\Journal#1#2#3#4{{#1} {\bf #2}, #3 (#4)}

\def\PLB{{\em Phys. Lett.}  B}

\def\PRep{\em Phys. Rep.}
\def\PRD{{\em Phys. Rev.} D}

\makeatother

\begin{document}

\thispagestyle{empty}

\begin{flushright}
\begin{tabular}{l}
{\tt LPT Orsay, 05-48}\\
{\tt PCCF RI 0504}\\
\end{tabular}
\end{flushright}
\vskip 2.6cm\par
\begin{center}
{\par\centering \textbf{\Large Lattice renormalization}}\\
\vskip .45cm\par
{\par\centering \textbf{\Large of the static quark derivative operator}}\\
\vskip .45cm\par
\vskip 0.9cm\par
{\par\centering 
\sc  B.~Blossier$^a$, 
 A.~Le~Yaouanc$^a$, V. Mor\'enas$^b$, O.~P\`ene$^a$}
{\par\centering \vskip 0.5 cm\par}
{\par\centering \textsl{$^a$ 
Laboratoire de Physique Th\'eorique (B\^at.210), Universit\'e
Paris XI-Sud,\\
Centre d'Orsay, 91405 Orsay-Cedex, France.} \\
\vskip 0.3cm\par}
{\par\centering \textsl{$^b$
Laboratoire de Physique Corpusculaire \\
Universit\'e Blaise Pascal - CNRS/IN2P3 F-63000 Aubi\`ere Cedex, France. } \\
}
\end{center}

\begin{abstract}
We give the analytical expressions and the numerical values of radiative corrections to
the covariant derivative operator on the static quark line, used for the lattice
calculation of the Isgur-Wise form factors $\tau_{1/2}(1)$ and $\tau_{3/2}(1)$. These corrections 
induce an enhancement of renormalized quantities if an hypercubic blocking procedure is used for
the Wilson line, while there is a reduction without such a procedure.
\end{abstract}
\vskip 2.2 cm

\setcounter{page}{1}
\setcounter{equation}{0}

\renewcommand{\thefootnote}{\arabic{footnote}}
\vspace*{-1.5cm}
\newpage
\setcounter{footnote}{0}

\section{Introduction}

In a previous paper \cite{IWOrsay} we proposed a method to compute 
on the lattice, in the static limit of HQET, the Isgur-Wise form factors 
$\tau_{1/2}(1)$ and $\tau_{3/2}(1)$ which 
parameterize decays of $B$ mesons into orbitally excited (P wave) 
$D^{\ast\ast}$ charmed mesons. Keep in mind that the zero recoil is the only definite 
limit of HQET 
on the lattice, because the Euclidean effective theory with a non vanishing spatial
momentum of the heavy quark is not lower bounded. Then it reveals impossible to calculate directly the 
$\tau_j$'s from the 
currents because the matrix elements vanish at zero recoil. To compute $\tau_{1/2}(1)$ and 
$\tau_{3/2}(1)$, we proposed to evaluate on the lattice the matrix elements  
$\langle H^\ast_0(v) | \bar h(v) \gamma_i\gamma_5 D_j h(v) | H(v)\rangle$ and
 $\langle H^\ast_2(v) |\bar h(v) \gamma_i\gamma_5 D_j h(v) | H(v)\rangle$, 
 using the following equalities:
 
\beq\label{scalaire}
\langle H^\ast_0(v) | \bar h(v) \gamma_i\gamma_5 D_j h(v) 
  | H(v)\rangle  =  i \, g_{ij}\left (M_{H^{\ast}_0} -
 M_{H}\right)  \tau_{\frac 1 2}(1), 
\eeq  
\beq\label{tenseur}
\langle H^\ast_2(v) |\bar h(v) \gamma_i\gamma_5 D_j h(v) 
  | H(v)\rangle  = - i \sqrt{3}\left (M_{H^{\ast}_2} - 
M_{H}\right)
\tau_{\frac 3 2}(1) \epsilon^\ast_{ij}\;,
\eeq
where $D_i$ is the covariant derivative ($D_i=\partial_i + i g A_i$), $M_{H}$, 
$M_{H^{\ast}_0}$ and $M_{H^{\ast}_2}$ are the mass of
the $0^-$, $0^+$ and $2^+$ states respectively, and $\epsilon^\ast_{ij}$ is the
polarization tensor. These relations are defined between renormalized quantities. Then 
we have to renormalize the matrix element of the derivative operator computed on the lattice. 
We explained that power and logarithmic divergences are not to be
feared in the zero recoil limit. However finite renormalization is present and we
 want to establish the one-loop contributions to the derivative operator with the hypercubic 
 blocking \cite{hasen2} of the Wilson line. 
 
We have to renormalize and to match onto the continuum the bare operator 
$O^B_{ij}=\bar{h}^B\gamma_i\gamma^5D_jh^B$, where $h^B$ is
the bare heavy quark field. We choose the MOM scheme whose renormalization conditions are: 
1) the renormalized heavy quark propagator is equal to the free one, and 2) the renormalized vertex 
function taken between renormalized external legs is the tree level one.

The rhs of
equations (\ref{scalaire}) and (\ref{tenseur}) 
are independent of the renormalization scale $\mu$. Indeed, on the one hand, 
$M_{H^{\ast}_0}-M_{H}\equiv \overline{\Lambda}_{0^+}-
 \overline{\Lambda}_{0^-}$ and $M_{H^{\ast}_2}-M_{H}\equiv \overline{\Lambda}_{2^+}-
 \overline{\Lambda}_{0^-}$ are physical quantities. On the other hand, 
 
\bea\nonumber
\frac{\lgl D_0|\bar{c}\gamma^\mu\gamma^5 b|B\rgl}{\sqrt{m_B m_{D_0}}}&\equiv& g^+ (v+v')^\mu +
g^-(v-v')^\mu\\
\nonumber
&\equiv& -\tau_{1/2}(\mu,w)\sqrt{w-1}F^\mu,\\
\nonumber
F^\mu&=&\sqrt{w+1}C^5_1(\mu,w)a^\mu + \sqrt{w-1}\left[C^5_2(\mu,w)
v^\mu+ C^5_3(\mu,w)v'^{\mu}\right],
\eea
where $\sqrt{w^2-1}\, a^\mu= (v-v')^\mu$, and the $C^5_i$'s are the matching coefficients between the QCD 
operator $\bar{c}\gamma^\mu \gamma^5b$ and the HQET operators $\bar{c}_{v'}\gamma^\mu \gamma^5b_v$, 
$\bar{c}_{v'}v^\mu \gamma^5b_v$ and $\bar{c}_{v'}v'^\mu \gamma^5b_v$ respectively.
\bea\label{matching}
\nonumber
\frac{\lgl D_0|\bar{c}\gamma^\mu\gamma^5 b|B\rgl}{\sqrt{w-1}\sqrt{m_B m_{D_0}}}
&=& \quad - \tau_{1/2} (\mu,w) F^\mu\\
&\longrightarrow_{w\to 1}& \quad - \tau_{1/2} (\mu,1) \sqrt{2}\, C^5_1(\mu,1)\, a^\mu.
\eea
$C^5_1(\mu,1)\equiv C^5_1(1)$ \cite{WiseManohar} ($C^5_1(1)\equiv \eta_A=0.986\pm 0.005$
 \cite{neubert}) and the lhs of (\ref{matching}) is also
independent of $\mu$: thus $\tau_{1/2}(\mu,1)\equiv \tau_{1/2}(1)$. We can use the same argument to 
prove 
the scale independence of $\tau_{3/2}(\mu,1)$. Consequently the scale $\mu$ will be omitted in the
following.

It is well known that the heavy quark self-energy diverges linearly in $1/a$ \cite{EichtenHill}, 
so we introduce a mass counterterm $\delta m$ to cancel this divergence. Numerically it is
canceled non-perturbatively in the ratio between the three-point function and the two-point 
functions to obtain a matrix element, or in the difference between binding energies of heavy-light 
mesons.

The bare heavy propagator on the lattice is
\bea\nonumber\label{propag}
S^B(p)&=&\frac{a}{1-e^{-ip_4a}+a\delta m + a\Sigma(p)}\\
\nonumber
&=&\frac{a}{1-e^{-ip_4a}} 
\sum_n\left(\frac{-a[\delta m+\Sigma(p)]}{1-e^{-ip_4a}}\right)^n\\
&\equiv&Z_{2h}S^R(p).
\eea
By choosing the renormalization conditions 
\beq\nonumber
(S^R)^{-1}(p)|_{ip_4\to 0}=ip_4, \quad \delta m=-\Sigma(p_4=0),
\eeq
the constant $Z_{2h}$ is then:
\beq\nonumber
Z_{2h}=1-\left.\frac{d\Sigma}{d(ip_4)}\right|_{ip_4\to 0}.
\eeq

The bare vertex function $V^B_{ij}(p)$ is defined as: 
\bea\nonumber\label{vertex}
V^B_{ij}(p)&=&(S^B)^{-1}(p)\sum_{x,y}e^{ip\cdot(x-y)}\lgl h^B(x)O^B_{ij}(0)\bar{h}^B(y)\rgl
(S^B)^{-1}(p)\\
&=&\frac{Z_{\cal D}}{Z_{2h}}(S^R)^{-1}(p)\sum_{x,y}e^{ip\cdot(x-y)}\lgl
h^R(x)O^R_{ij}(0)\bar{h}^R(y)\rgl(S^R)^{-1}(p), 
\eea
where
\beq\nonumber
O^B_{ij}(0)=Z_{\cal D}\, O^R_{ij}(0).
\eeq

We will see below that $V^B_{ij}(p)$ can be written as 
\bea\nonumber
V^B_{ij}(p)\nonumber
&=&(1+\delta V)\bar{u}(p)\gamma_i\gamma^5p_ju(p)\\
&\equiv&(1+\delta V)V^R_{ij}(p).
\eea
$\delta V$ is given by all the 1PI one-loop diagrams containing the vertex.



We obtain $\lgl H^{**}|O^R_{ij}|H\rgl=
Z^{-1}_{\cal D}\lgl H^{**}|O^B_{ij}|H\rgl$ where 
$Z_{\cal D}=Z_{2h}(1+\delta V)$ and $\lgl H^{**}|O^B_{ij}|H\rgl$ was computed 
on the lattice.

This paper is organized as follows: in Sec. \ref{sec2} we recall the action we use
for the heavy quark, we give the corresponding Feynman rules and we clarify our notations; 
in Sec. \ref{sec3} we give the analytical expression for the heavy quark self-energy, 
in Sec. \ref{sec4}
 we give the analytical expression for radiative corrections to the derivative operator in 
 lattice HQET. We briefly conclude in Sec. \ref{sec5}.

\section{Heavy quark action and Feynman rules}\label{sec2}

The lattice HQET action for the static heavy quark is 

\beq\label{actionHQET}
S^{\rm HQET} = a^3\sum_n \left\{h^{\dag}(n)\left[h(n)-U^{\dag,\rm HYP}_4(n-\hat{4})
h(n-\hat{4})\right]+a\delta m h^{\dag}(n)h(n)\right\},\\
\eeq
where $U^{\rm HYP}_4(n)$ is a link built from an hypercubic blocking. 

We will use in the rest of the paper the following notations taken from 
\cite{Capitani}-\cite{Sharpe}: 
\beq\nonumber
\int_p \equiv \int_{-\frac{\pi}{a}}^{\frac{\pi}{a}} \frac{d^4p}{(2\pi)^4}\;,\quad\quad
\int_{\vec{p}} \equiv \int_{-\frac{\pi}{a}}^{\frac{\pi}{a}} \frac{d^3p}{(2\pi)^3}\;,
\quad\quad a^4 \sum_n e^{ipn} = \delta(p),
\eeq
\beq\nonumber
\int_k \equiv \int_{-\pi}^{\pi} \frac{d^4k}{(2\pi)^4}\;,\quad\quad
\int_{\vec{k}} \equiv \int_{-\pi}^{\pi} \frac{d^3k}{(2\pi)^3}\;,
\eeq
\beq\nonumber
h(n)=\int_p e^{ipn}h(p),
\eeq
\beq\nonumber
U_\mu(n)=e^{iag_0A^a_\mu(n)T^a}=1+iag_0A^a_\mu(n)T^a-\frac{a^2g_0^2}{2!}A^a_\mu(n)A^b_\mu(n)T^aT^b
+{\cal O}(g^3_0),
\eeq
\beq\nonumber
U^{\rm HYP}_\mu(n)=e^{iag_0B^a_\mu(n)T^a}=1+iag_0B^a_\mu(n)T^a-\frac{a^2g^2}{2!}B^a_\mu(n)B^b_\mu(n)T^aT^b
+{\cal O}(g^3_0),
\eeq
\beq\nonumber
A^a_\mu(n)=\int_p e^{ip(n+\frac{a}{2})}A^a_\mu(p), \quad\quad
B^a_\mu(n)=\int_p e^{ip(n+\frac{a}{2})}B^a_\mu(p),
\eeq
\beq\nonumber
\Gamma_{\lambda}=\sin ak_{\lambda},
\eeq
\beq\nonumber
c_\mu=\cos \left(\frac{a(p+p')_\mu}{2}\right), \quad\quad
s_\mu=\sin \left(\frac{a(p+p')_\mu}{2}\right),
\eeq
\beq\nonumber
M_\mu=\cos \left(\frac{k_\mu}{2}\right), \quad\quad
N_\mu=\sin \left(\frac{k_\mu}{2}\right),
\eeq
\beq\nonumber
W=2\sum_{\lambda}\sin^2 \left(\frac{k_\lambda}{2}\right).
\eeq

In the Fourier space, the action is given at the order of ${\cal O}(g^2_0)$ by:
\bea\label{actionfourier}\nonumber
S^{\rm HQET}&=&\int_p a^{-1}h^{\dag}(p)(1-e^{-ip_4a})h(p) + \delta m h^{\dag}(p)h(p)\\
\nonumber
&+&ig_0\int_p\int_{p'}\int_q \delta(q+p'-p) (h^{\dag}(p)B^a_4(q)T^ah(p')e^{-i(p_4+p'_4)\frac{a}{2}}\\
&+&\frac{ag^2_0}{2!}\int_p\int_{p'}\int_q\int_r \delta(q+r+p'-p)
h^{\dag}(p)B^a_4(q)B^b_4(r)T^aT^bh(p')e^{-i(p_4+p'_4)\frac{a}{2}}. 
\eea
The block gauge fields $B^a_\mu$ can be expressed in terms of the usual gauge fields:

\beq\nonumber
B_\mu=\sum_{n=1}^{\infty} B^{(n)}_\mu,
\eeq
where $B^{(n)}_\mu$ contains $n$ factors of $A$. At NLO, it was shown that we only need 
 $B^{(1)}_\mu$ \cite{Lee}:
 
\bea\nonumber
B^{(1)}_\mu(k)&=&\sum_{\nu} h_{\mu\nu}(k)A_\nu(k),\\
\nonumber
h_{\mu\nu}(k)&=&\delta_{\mu\nu}D_\mu(k) + (1-\delta_{\mu\nu})G_{\mu\nu}(k),\\
\nonumber
D_\mu(k)&=&1-d_1\sum_{\rho \neq \mu} N^2_\rho +d_2 \sum_{\rho < \sigma, \rho,\sigma \neq \mu}
N^2_\rho N^2_\sigma -d_3 N^2_\rho N^2_\sigma N^2_\tau,\\
\nonumber
G_{\mu\nu}(k)&=&N_\mu N_\nu \left(d_1-d_2 \frac{N^2_\rho + N^2_\sigma}{2} + d_3 
\frac{N^2_\rho N^2_\sigma}{3}\right),\\
\nonumber
\eea
\beq\nonumber
d_1=(2/3)\alpha_1(1+\alpha_2(1+\alpha_3)), \quad\quad d_2=(4/3) \alpha_1\alpha_2(1+2\alpha_3),
\quad\quad d_3=8\alpha_1\alpha_2\alpha_3.
\eeq
Two sets of $\alpha_i$'s have been chosen: 1) $\alpha_1=0.75, 
\alpha_2=0.6, \alpha_3=0.3$ (which has been chosen in our simulation and has been motivated
in \cite{hasen2}) and 2) $\alpha_1=1.0, \alpha_2=1.0, \alpha_3=0.5$,
motivated in \cite{Alpha2005}. We will label these two sets respectively by HYP1 and HYP2.

\vspace{0.7cm}
The Feynman rules can be easily deduced (they must be completed by the application of $h_{\mu\nu}$):

\vspace{0.5cm}

\begin{tabular}{|c|c|}
\hline
heavy quark propagator& $a(1-e^{-ip_4a}+\epsilon)^{-1}$\\
vertex $V^a_{\mu,hhg}(p,p')$&$-ig_0 T^a \delta_{\mu4} 
e^{-i(p_4+p'_4)\frac{a}{2}}$\\
vertex $V^{ab}_{\mu\nu,hhgg}(p,p')$&$-\frac{1}{2}ag^2_0 \delta_{\mu4}
\{T^a,T^b\}e^{-i(p_4+p'_4)\frac{a}{2}}$\\
gluon propagator in the Feynman gauge&$a^2\delta_{\mu\nu}\delta^{ab}(2W + a^2\lambda^2)^{-1}$\\
\hline
\end{tabular}

\vspace{0.5cm}

Note that $p'$ and $p$ are the in-going and the out-going fermion momenta, respectively.
We also introduce an infrared regulator $\lambda$ for the gluon propagator. We symmetrize the vertex 
$V^{ab}_{\mu\nu,hhgg}$ by introducing the anti-commutator of the $SU(3)$ generators, normalized by a 
factor $\frac 1 2$. The gluon propagator and the vertices are defined with the $A$ field. 
The coefficient $\sum_{i=1}^3 h_{4i}^2\equiv H(N_4)$ will enter as a global multiplicative factor in all 
the integrals expressed below. We have chosen the Feynman gauge: since one calculates the
renormalization of a gauge-invariant operator, the renormalization factor $Z_{\cal D}$ is
gauge invariant.

\section{Heavy quark self-energy}\label{sec3}

From (\ref{propag}) we have $\Sigma(p) =-(F_1+F_2)$, 
where $F_1$ and $F_2$ correspond to the diagrams shown in Fig. \ref{self}(a) and (b):

\begin{figure}[b]

\begin{center}

\begin{tabular}{ccc}

\begin{picture}(-10,30)(10,30)

\Line(-45,30)(55,30)
\Line(-45,31)(55,31)
\CTri(-23,27)(-23,34)(-18,30.5){Black}{Black}
\CTri(2,27)(2,34)(7,30.5){Black}{Black}
\CTri(27,27)(27,34)(32,30.5){Black}{Black}
\GlueArc(5,30.5)(12.5,0,180){1}{10}
\Text(30,20)[c]{\small{$p$}}
\Text(5,20)[c]{\small{$p$+$k$}}
\Text(-20,20)[c]{\small{$p$}}

\end{picture}

&\rule[0cm]{2.0cm}{0cm}&
\begin{picture}(-10,30)(10,30)

\Line(-35,30)(45,30)
\Line(-35,31)(45,31)
\CTri(-18,27)(-18,34)(-13,30.5){Black}{Black}
\CTri(22,27)(22,34)(27,30.5){Black}{Black}
\GlueArc(5,44)(12.5,0,360){1}{20}
\Text(25,20)[c]{\small{$p$}}
\Text(-15,20)[c]{\small{$p$}}

\end{picture}
\\
\\
(a): Sunset diagram&\rule[0cm]{2.0cm}{0cm}&
(b):  Tadpole diagram\\
\end{tabular}
\end{center}
\caption{\label{self}: Self-energy corrections}
\end{figure}

\bea\label{diag1}\nonumber
F_1&=&-\frac{4}{3a}g^2_0 \int_k 
\frac{H(N_4)}{2W+a^2\lambda^2}\frac{e^{-i(k_4+2ap_4)}}{1-e^{-i(k_4+ap_4)}+ \epsilon}\\
\nonumber
&=&-\frac{1}{3a}g^2_0 \int_k \frac{H(N_4)}{N_4^2+E^2}
\frac{e^{-i(k_4+2ap_4)}}{1-e^{-i(k_4+ap_4)}+ \epsilon}\\
\nonumber
&=&-\frac{1}{3a}g^2_0 \int_k \frac{H(N_4)}{(N_4+iE)(N_4-iE)}
\frac{e^{-i(k_4+2ap_4)}}{1-e^{-i(k_4+ap_4)}+ \epsilon}\\
&=&-\frac{1}{3a}g^2_0 \int_{\vec{k}} \frac{1}{E}\frac{H(-iE)}{\sqrt{1+E^2}}
\frac{e^{-2iap_4}}{e^{E'}-e^{-iap_4}}.
\eea
Note that Latin indices are spatial and
 
\beq\nonumber
E^2=\sum_iN^2_i + \frac{a^2\lambda^2}{4},
\eeq
\bea\nonumber 
H(N_4)&=&\left(1-d_1\sum_i N^2_i +d_2\sum_{i<j}N^2_iN^2_j-d_3 N^2_1N^2_2N^2_3\right)^2\\
\nonumber
&+&N^2_4\sum_iN^2_i\left(d_1-\frac{d_2}{2} \sum_{j\ne i}N^2_j + \frac{d_3}{3} 
\prod_{j\ne i} N^2_j\right)^2,  
\eea
\beq\nonumber
E'=2 \mbox{argsh} (E).
\eeq

In (\ref{diag1}) we have eliminated properly the non covariant pole 
$k_4=-ap_4+i\epsilon$ by closing the integration contour in the complex plane $\Im (k_4)<0$ where
there is the single pole $N_4=-iE$. Furthermore the integrals along the lines $k_4=\pm \pi + 
ik'_4$ are equal, because the integrand is $2 \pi$-periodic.

Finally we have in the limit $ap_4\to 0$:

\bea\nonumber
F_1&=&\frac{4}{3a}g^2_0\int_{\vec{k}} \frac{H(-iE)}{4E\sqrt{1+E^2}}\frac{1}{1-e^{E'}}\\
&+&\frac{4}{3}g^2_0 i p_4 \int_{\vec{k}} \frac{H(-iE)}{2E\sqrt{1+E^2}}
\left[\frac{1}{e^{E'}-1}+\frac 1 2 \frac{1}{(e^{E'}-1)^2}\right].
\eea

We find:

\beq\label{f1}
F_1\equiv-\frac{g^2_0}{12\pi^2}\left\{f_1(\alpha_i)/a+ip_4[2\,\mbox{ln}(a^2\lambda^2)+f_2(\alpha_i)]
\right\}.
\eeq

The tadpole diagram $F_2$ is 
\bea\label{f2}
\nonumber
F_2&=&-\frac{1}{2}\frac{4g^2_0}{3a}\,e^{-iap_4}\int_k \frac{H(N_4)}{2W}\\
\nonumber
&=_{ap_4\to 0}&-\frac{1}{2}\frac{4g^2_0}{3}(1/a-ip_4)\int_k \frac{H(N_4)}{2W}\\
&\equiv&-\frac{g^2_0}{12\pi^2}(1/a -ip_4)\, f_3(\alpha_i)\, .
\eea
The factor 1/2 is introduced to compensate the over-counting of the factor 2 in the Feynman 
rule of the 2-gluon vertex when a closed gluonic loop is computed.
We can point out that the divergent part 
\beq\label{selfenergie}
\Sigma_0(\alpha_i)=\frac{g^2_0}{12\pi^2 a}\, \sigma_0(\alpha_i), \quad
\sigma_0=f_1+f_3
\eeq
of the self-energy is smaller with the sets HYP1 and HYP2 of the $\alpha_i$'s than with the 
corresponding contribution without "hypercubic" links 
\cite{EichtenHill}, as shown in the Table \ref{tab1}: $\sigma_0\,(\alpha_i=0)\,=19.95$,
$\sigma_0\,(\rm{HYP1})\,=5.76$ and $\sigma_0\,(\rm{HYP2})\,=4.20$, in good agreement with computations 
made by the ALPHA collaboration \cite{Alpha2005}, which compares the pseudoscalar heavy-light 
meson effective energy with different static heavy quark actions, and by Hasenfratz {\it et al}
\cite{hasen}. Qualitatively, one expects that the hypercubic blocking reduces the radiative 
corrections since it amounts roughly to introduce an additional cut-off in the integrals.

The wave function renormalization $Z_{2h}$ is 
\beq\label{Z_h}
Z_{2h}(\alpha_i)=1+\frac{g^2_0}{12\pi^2} \left[-2\,\mbox{ln}(a^2\lambda^2)+z_2(\alpha_i)\right], \quad
z_2=f_3-f_2.
\eeq
$|z_2|$ is also reduced by the hypercubic blocking, as indicated on Table \ref{tab1}:
$z_2\,(\alpha_i=0)\,=24.48$ \cite{EichtenHill}, $z_2\,(\rm{HYP1})\,=2.52$ and 
$z_2\,(\rm{HYP2})\,=-3.62$.

\section{Derivative operator in lattice HQET}\label{sec4}

We have to renormalize the operator $O^B_{ij}=\bar{h}^B\gamma_i\gamma^5D_jh^B$. Following
\cite{Capitani},

\bea\label{derivee} \nonumber
a^4 \sum_nO^B_{ij}(n)&=&a^4 \frac{1}{2a}\sum_n\bar{h}^B(n)\gamma_i \gamma^5 
U_j(n)h^B(n+\hat{j} )
-\bar{h}^B(n)\gamma_i \gamma^5U^{\dag}_j(n-\hat{j} )h^B(n-\hat{j} )\\
\nonumber
&=&\int_p\int_p' a^{-1}\delta(p-p')\bar{h}^B(p) \left(i \gamma_i \gamma^5 s_j \right)h^B(p')\\
\nonumber
&+&ig_0\int_p\int_{p'}\int_q \delta(q+p'-p) \bar{h}^B(p) \gamma_i \gamma^5 c_i 
A^a_j(q)T^ah^B(p')\\
&-&\frac{iag^2_0}{2!}\int_p\int_{p'}\int_q\int_r \delta(q+r+p'-p) 
\bar{h}^B(p) T^aT^b \gamma_i \gamma^5 s_j A^a_j(q)A^b_j(r) h^B(p').  \hspace{1cm} 
\eea

Note that we have chosen to not submit the covariant derivative to the hypercubic blocking. 
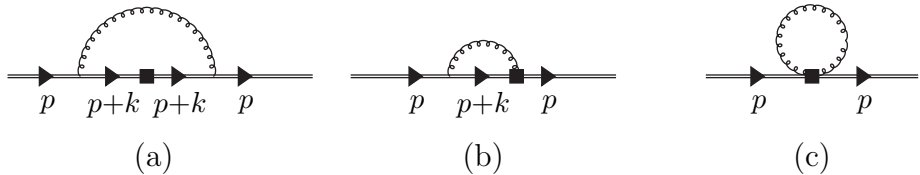
\begin{figure}[b]

\begin{center}

\begin{tabular}{ccccc}

\begin{picture}(-10,10)(10,40)

\Line(-50,30)(65,30)
\Line(-50,31)(65,31)
\CTri(-38,27)(-38,34)(-33,30.5){Black}{Black}
\CTri(-13,27)(-13,34)(-8,30.5){Black}{Black}
\CTri(12,27)(12,34)(17,30.5){Black}{Black}
\CTri(37,27)(37,34)(42,30.5){Black}{Black}
\GlueArc(2.5,30.5)(25,0,180){1}{20}
\Text(40,20)[c]{\small{$p$}}
\Text(15,20)[c]{\small{$p$+$k$}}
\Text(-10,20)[c]{\small{$p$+$k$}}
\Text(-35,20)[c]{\small{$p$}}
\CBox(0,28)(5,33){Black}{Black}
\end{picture}

&\rule[0cm]{3.0cm}{0cm}&
\begin{picture}(-10,10)(10,40)

\Line(-45,30)(55,30)
\Line(-45,31)(55,31)
\CTri(-23,27)(-23,34)(-18,30.5){Black}{Black}
\CTri(2,27)(2,34)(7,30.5){Black}{Black}
\CTri(27,27)(27,34)(32,30.5){Black}{Black}
\GlueArc(5,30.5)(12.5,0,180){1}{10}
\Text(30,20)[c]{\small{$p$}}
\Text(5,20)[c]{\small{$p$+$k$}}
\Text(-20,20)[c]{\small{$p$}}
\CBox(15,28)(20,33){Black}{Black}

\end{picture}

&\rule[0cm]{3.0cm}{0cm}&
\begin{picture}(-10,10)(10,40)

\Line(-35,30)(45,30)
\Line(-35,31)(45,31)
\CTri(-18,27)(-18,34)(-13,30.5){Black}{Black}
\CTri(22,27)(22,34)(27,30.5){Black}{Black}
\GlueArc(5,44)(12.5,0,360){1}{20}
\Text(25,20)[c]{\small{$p$}}
\Text(-15,20)[c]{\small{$p$}}
\CBox(2.5,28)(7.5,33){Black}{Black}

\end{picture}
\\
\vspace{0.5cm}
&&&&\\
(a)&\rule[0cm]{3.0cm}{0cm}&(b)&\rule[0cm]{3.0cm}{0cm}&(c)\\

\end{tabular}
\end{center}
\caption{\label{opder}: Operator corrections}
\end{figure}

The vertex function $V^B_{ij}$ is obtained by writing $V^B_{ij}=V^0_{ij}+V^1_{ij}+V^2_{ij}$, 
corresponding to the diagrams (a), (b) and (c) in Fig. \ref{opder}; 
$V^k_{ij}(\alpha_i)=\bar{u}(p)\gamma_i \gamma^5 u(p)V^k_j(\alpha_i)$, $k=0$, 1, 2.
The contribution $V^0_{ij}$ is then given by computing 
\bea\nonumber
V^0_j(\alpha_i)&=&-\frac{4i}{3a}g^2_0 \int_k \frac{H(N_4)}{2W+a^2\lambda^2} \sin\left(k+ap\right)_j
\frac{e^{-i(k_4+2ap_4)}}{(1-e^{-i(k_4+ap_4)}+\epsilon)^2}\\
\nonumber
&=&-\frac{4i}{3a}g^2_0 \int_k \frac{H(N_4)}{2W+a^2\lambda^2} \left(\Gamma_j + ap_j \cos k_j\right)
e^{-iap_4}\left(\frac{e^{-i\frac{k_4+ap_4}{2}}}{1-e^{-i(k_4+ap_4)}+\epsilon}\right)^2\\
\nonumber
&=&-\frac{4i}{3a}g^2_0 \int_k \frac{H(N_4)}{2W+a^2\lambda^2} \left(\Gamma_j + ap_j \cos k_j\right)
(1-iap_4)\frac{1}{\left[2i\sin \left(\frac{k_4+ap_4}{2}\right)+e^{i\frac{k_4+ap_4}{2}}\epsilon
\right]^2}.\\
\eea
We can get rid of the integrand proportional to $\Gamma_j$, because it is an odd 
term. It remains
\bea\nonumber
V^0_j(\alpha_i)&=&-\frac 4 3 ig^2_0p_j \int_k \frac{H(N_4)}{2W+a^2\lambda^2} 
\frac{\cos k_j}{(2iN_4+\epsilon M_4)^2}.
\eea
The integrand has poles at $N_4=\pm iE$, $k_4=2i \mbox{argth}\left(\frac
{\epsilon}{2}\right)$. Once again we close the integration contour around the single pole 
$N_4=- iE$. 
Since $j$ is spatial, the "sail" diagram drawn on Fig. 
\ref{opder}(b) does not give any contribution, thus $V^1_{ij}=0$. There is finally the tadpole 
contribution $V^2_{ij}$ which is given by computing
\bea
V^2_j(\alpha_i)&=&-\frac{1}{2!} \frac 4 3 ig^2_0 p_j \int_k \frac{1}{2W} = 
-\frac{ig^2_0}{12\pi^2} p_j\, 12.23.
\eea

We have finally $\lgl H^{**}|O^R_{ij}|H\rgl = Z^{-1}_{\cal D}(\alpha_i)\lgl
H^{**}|O^B_{ij}|H\rgl(\alpha_i) $
where 
\bea\label{ZIW}\nonumber
Z_{\cal D}(\alpha_i)&=&Z_{2h}(\alpha_i)[1+\delta V(\alpha_i)],\\
\nonumber
\delta V(\alpha_i)&=&-\frac{4g^2_0}{3} \int_k \left(\frac{H(N_4)\cos k_j}{(2W+a^2\lambda^2)
(2iN_4+\epsilon M_4)^2} + 
\frac{1}{4W}\right)\\
\nonumber
&=&-\frac{4g^2_0}{3}\frac{1}{2}\int_{\vec{k}}\frac{H(-iE)\cos k_j}{(2E)^3\sqrt{1+E^2}}
-\frac{g^2_0}{12\pi^2}12.23\\
&\equiv&\frac{g^2_0}{12\pi^2} \left[2\,\mbox{ln}(a^2\lambda^2)+f_4(\alpha_i)\right],
\eea
\beq\label{fin}
Z_{\cal D}(\alpha_i) = 1+\frac{g^2_0}{12\pi^2} z_d(\alpha_i), \quad  z_d=z_2+f_4.
\eeq
The numerical values of $z_d$ are indicated in Table \ref{tab1}.


\begin{table}
\begin{center}
\begin{tabular}{|c|c|c|c|}
\hline
&$\alpha_i=0$&HYP1&HYP2\\
\hline
$f_1$&7.72&1.64&-1.76\\
$f_2$&-12,25&1.60&9.58\\
$f_3$&12.23&4.12&5.96\\
$f_4$&-12.68&-10.32&-8.18\\
\hline
$\sigma_0$&19.95&5.76&4.20\\
$z_2$&24.48&2.52&-3.62\\
$z_d$&11.80&-7.80&-11.80\\
\hline
\end{tabular}
\end{center}
\caption{\label{tab1} Numerical values of the parameters $f_1$, $f_2$, $f_3$, $f_4$, 
$\sigma_0$, $z_2$, $z_d$ defined in equations (\ref{f1}), (\ref{f2}), (\ref{ZIW}), (\ref{selfenergie}), 
(\ref{Z_h}) and (\ref{fin}), respectively.}
\end{table}

Remark that infrared divergences appearing in $Z_{2h}$ and $1+\delta V$ cancel and there is no 
dependence on $a$,
a further consequence of 
the $\mu$ independence of the matrix element 
$\langle H^\ast_0(v') | \bar h(v') \gamma_i\gamma_5 D_j h(v) | H(v)\rangle$ at zero recoil. 
Note also that as already mentioned the tadpole 
diagram of the operator vertex is not smoothed by the hypercubic blocking; therefore it is quite 
large. On the
other hand the tadpole contribution to the self-energy is smoothed by the blocking. The final 
result is a large positive correction to the renormalized matrix element. By fixing $g_0=1$ 
($\beta=6.0$), this gives $Z^{-1}_{D}(\rm{HYP1}) = 1.07$ and $Z^{-1}_{D}(\rm{HYP2}) = 1.10$, thus 
the discrepancy between $Z^{-1}_{D}(\rm{HYP1})$ and $Z^{-1}_{D}(\rm{HYP2})$ is small (3\%); 
without hypercubic 
blocking one would obtain $Z^{-1}_{D}(\alpha_i=0) = 0.90$. We can think of applying a boosting 
procedure; in the pure HYP case the boosting plaquette correction is very small \cite{Sharpe} 
(equation 19 and below). On the other hand we have to take into
account that the covariant derivative operator involves links without hypercubic blocking, 
therefore one should employ a different prescription for the diagram drawn on Fig. 
\ref{opder}(c), possibly
leading to a larger tadpole contribution from the operator to $Z^{-1}_{D}(\alpha_i\neq 0)$, 
and therefore a larger positive correction. Anyhow this kind of recipe would not lead to 
$Z^{-1}_{D}(\alpha_i\neq 0)$ significantly larger than 1.1.

With our exploratory lattice study and taking account $Z^{-1}_{\cal D}(\rm{HYP1})$, we find 
$\taud(1)=0.41(5)(?)$, $\taut(1)=0.57(10)(?)$ and $\taut^2(1)-\taud^2(1)=0.15(10)$, where 
systematics are unknown; one is then not too far (within 1$\sigma$) from saturating by 
ground states the Uraltsev sum rule 
\cite{Uraltsev} $\sum_n|\taut^{(n)}(1)|^2 - |\taud^{(n)}(1)|^2 = \frac 1 4$. However the relation 
$\mu^2_{\pi}-\mu^2_G > 9 \, \Delta^2 \, \taud^2(1)$ \cite{ineqmom}, with 
$\Delta\equiv M_{H^*_0}-M_H=0.4$ GeV \cite{IWOrsay}, 
$\mu^2_G=0.35$ ${\rm GeV}^2$, leads to $\mu^2_\pi$ larger than 0.6 ${\rm GeV}^2$, which is 
significantly above experimental determination by moments.

\section{Conclusion}\label{sec5}

In this paper we have calculated the radiative corrections to the covariant derivative operator 
$\bar{h}\gamma_i\gamma^5D_jh$ in lattice HQET with an hypercubic blocking of the Wilson line
defining the heavy quark propagator. This determines the renormalization of the operator which is used
to estimate the Isgur-Wise functions between the ground state and the $L=1$ excitations at zero recoil.
We find that there is a global, but moderate, enhancement of $\tau_{\frac 1 2}(1)$ and 
$\tau_{\frac 3 2}(1)$ with 
respect to the bare quantities computed on the lattice in the case where one introduces fat 
timelike links, while there is a reduction with a simple Wilson line.

\end{document}